\documentclass[aps,pra,twocolumn,superscriptaddress,amsmath,showpacs]{revtex4}
\usepackage{graphicx}
\usepackage{bm}
\usepackage{times}
\begin{document}
\title{Charge oscillation-induced light transmission through 
subwavelength slits and holes}
\author{X.~R.~Huang} 
\email[]{xrhuang@aps.anl.gov}
\affiliation{Advanced Photon Source, Argonne National Laboratory, 
Argonne, Illinois 60439, USA}
\author{R.~W. Peng} 
\affiliation{National Laboratory of Solid State Microstructures, 
Nanjing University, Nanjing 210093, China}
\author{Z.~Wang} 
\affiliation{National Laboratory of Solid State Microstructures, 
Nanjing University, Nanjing 210093, China}
\author{F.~Gao} 
\affiliation{National Laboratory of Solid State Microstructures, 
Nanjing University, Nanjing 210093, China}
\author{S.~S.~Jiang} 
\affiliation{National Laboratory of Solid State Microstructures, 
Nanjing University, Nanjing 210093, China}

\begin{abstract}
We present a concrete picture of spoof surface plasmons (SSPs)
combined with cavity resonance to clarify the basic 
mechanism underlying extraordinary light transmission through metal films
with subwavelength slits or holes.
This picture may indicate a general mechanism of metallic nanostructure optics: 
When light is incident on a non-planar conducting surface, the free electrons 
cannot move homogeneously in response to the incident electric field, 
i.e., their movement can be impeded at the rough parts, 
forming inhomogeneous charge distributions.  
The oscillating charges/dipoles then emit photons (similar to
Thomson scattering of x rays by oscillating electrons), and the interference between the photons
may give rise to anomalous transmission, reflection or scattering.
\end{abstract}
\pacs{42.25.Bs, 42.79.Dj, 78.67.-n, 84.40.Az}
\maketitle

Since the discovery of extraordinary light transmission 
through metal films perforated by subwavelength hole arrays \cite{r1}, 
tremendous theoretical and experimental work has been carried 
out to understand the underlying physics. Several mechanisms, 
particularly surface plasmons (SPs), have been proposed as the possible 
origins \cite{r2,r3,r4,r5,r6,r7,r8}.  
However, no universal understanding
has been reached to date. Here based on first-principle calculations, 
we present a charge oscillation-induced light emission mechanism, 
which gives the origin of enhanced transmission in subwavelength 
systems and 
may also shed 
light on the fundamental of 
interactions between light and metallic nanostructures.

In non-magnetic media, the electric and 
magnetic fields are coupled by Maxwell's equations 
$\nabla \times {\mathbf E} = -iK{\mathbf H}$
and $\nabla \times {\mathbf H} = iK\varepsilon{\mathbf E}$ 
($K = 2\pi/\lambda$, $\lambda$ the wavelength in vacuum, and 
$\varepsilon$ the effective permittivity). 
For varying $\varepsilon({\mathbf r}$) (i.e., $\nabla\varepsilon\neq 0$), 
the divergence of the second equation generally gives 
$\nabla \cdot {\mathbf E} = -[(\nabla \varepsilon) 
\cdot {\mathbf E}]/\varepsilon = 4\pi \rho \neq 0$, 
where $\rho$ is the charge density. 
Consider in Fig.~1 the free-standing one-dimensional (1D) gold grating 
illuminated by a plane wave, where Maxwell's equations can be solved by 
the rigorous coupled-wave analysis (RCWA), a first-principle method  
\cite{r9}. 
For simplicity, we only discuss normal incidence in this Letter. 
Figure 1 shows the zero-order transmittance 
($T_0$) spectra of both TM (${\mathbf H} \, \| \, \hat{\mathbf y}$) \cite{r4,r5} 
and TE (${\mathbf E} \, \| \, \hat{\mathbf y}$) waves \cite{r10}.
Here RCWA correctly reveals the cutoff wavelength $\lambda_c \simeq 2W$, 
above which transmission of TE waves is forbidden. 
The reason is that TE waves in the slit approximately 
take the waveguide modes 
$E_y \propto \sin(m\pi x/W )\exp[\pm \, i\pi z (4 /\lambda^2 - m^2/W^2)^{1/2}]$
($m \neq 0$ being integers) \cite{r11},
where for $\lambda > 2W$, all the
modes are evanescent.
The drastic difference between the TE and TM spectra stems 
from the fact that TE waves satisfy $\nabla \cdot {\mathbf E} \equiv 0$
while TM waves may induce electric charge oscillation 
$\rho({\mathbf r})e^{i\omega t}$ ($\omega$ the frequency), where
$\rho({\mathbf r}) = \nabla \cdot {\mathbf E}({\mathbf r})/4\pi$.

\begin{figure}[b] 
\includegraphics[scale=1.0,angle=0]{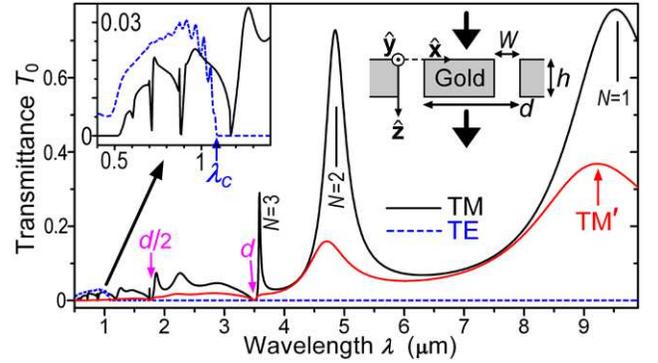}
\caption{(color online) Transmission spectra of a gold grating with 
period $d = 3.5$, slit width $W = 0.5$, and thickness 
$h = 4$ $\mu$m. $\hat{\mathbf x}$, 
$\hat{\mathbf y}$, $\hat{\mathbf z}$ are unit vectors.  
The TM$'$ curve was calculated with Re[$\varepsilon_c(\omega)$] 
($< 0$) for gold replaced by $-$Re[$\varepsilon_c(\omega)$].}
\end{figure}

For TM polarization, the vertical component $E_z(x,z)$ of the electric field 
(invariant with $y$) 
has abrupt discontinuity across the surfaces at $z_s=0$ and $h$,
from which one obtains the \emph{surface charge density\/} 
$\tilde{\rho}_s(x, z_s) = \delta E_z(x, z_s)/4\pi$. 
In Fig.~2(a), the calculated $\tilde{\rho}_s(x, 0)$ profile 
correctly shows that charges only exist on the metal surface
[$\tilde{\rho}_s(x,0)=0$ for $0\!<\!x\!<\!W$], and 
the charges tend to accumulate at the metal corners. 
At resonant wavelengths,
$\tilde{\rho}_s(x,0)\simeq (-1)^N \tilde{\rho}_s(x,h)$, 
i.e., the charge patterns on the two surfaces are nearly the same,
but they have opposite signs for odd resonant numbers 
$N$ (defined below).
The \emph{bulk charge density\/} is given by 
$\rho_v = \nabla \cdot {\mathbf E}/4\pi$. 
In Fig.~2(b), the $\rho_v (x, z = \mbox{const})$ curve calculated from 
the internal ${\mathbf E}$ eigenmodes reveals 
that there is a strong peak exactly centered at each 
slit wall, $x_w = 0$ or $W$ (plus any multiple of $d$). 
When sufficient diffraction orders (1601 orders in Fig.~2) are retained in RCWA,
$\rho_v (x, z = \mbox{const})$ approaches a delta function across $x_w$
[see Inset of Fig.~2(b)],
i.e. charges inside the grating also appear as surface charges on the walls.

\begin{figure} 
\includegraphics[scale=1.0,angle=0]{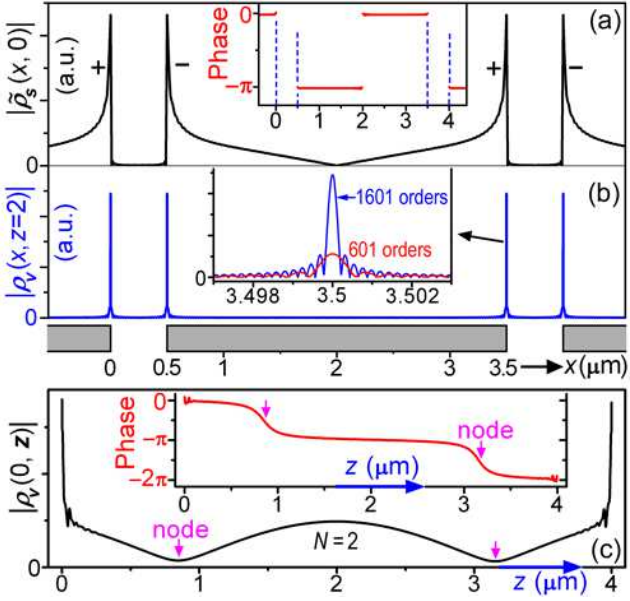} 
\caption{(Color online) Charge distributions in the gold grating of Fig.~1 under resonance.
TM polarization. $\lambda = 4.845$ $\mu$m (for peak $N=2$ in Fig.~1).
(a) Surface charge densities. (b) Sectional charge densities. 
(c) Charge densities on the wall.}
\end{figure}

The profile $\rho_v(0, z)$ [$\equiv -\rho_v(W,z)$] 
plotted in Fig.~2(c) shows that in the central range the charge density on the wall 
is nearly a standing wave (with wavevector $k_z \simeq 2\pi  /\lambda$).
This indicates that the electric 
field in the slit is also a standing wave \cite{r12,r13} consisting of a forward wave
${\mathbf E}_a \simeq E_a \exp(-ik_z z) \hat{\mathbf x}$ and a backward wave 
${\mathbf E}_b \simeq E_b \exp(ik_z z) \hat{\mathbf x}$ in Fig.~3. 
Note that $|\rho_v(x_w,z)|$
increases sharply when $z\rightarrow z_s$ in Figs.~2(c).
Mathematically, the charge density at each corner consists 
of both $\tilde{\rho}_s(x_w,z_s)$ and $\rho_v(x_w, z\! \rightarrow \! z_s)$.
Since they are always in phase, these two contributions are superimposed constructively.
This makes the total charge densities at the corners much higher than
in other regions, which is a typical edge effect. 
Consequently, there exist two large oscillating dipoles at the two 
ends of the slit. 

Now we may obtain a clear picture about the light scattering process. 
As shown in Fig.~3, the incident electric field ${\mathbf E}_{in}$ drives 
(mainly) free electrons on the metal surface to move, but the
movement is blocked at one corner of the metal, resulting in accumulation of electrons there. 
Meanwhile, extra positive charges appear at the other corner since some of the electrons
have moved away. Thus, two dipoles 
${\mathbf P}_a$ and ${\mathbf P}_r$ are formed. Since they oscillate with
the incident wave, these dipoles act as \emph{light sources\/}
emitting new wavelets (photons), which form scattered waves 
(Thomson scattering). 
If we consider each period of the array as an overall scattering unit, the wavelets
emitted from two adjacent units have a path difference $d\sin\theta$ along an arbitrary 
direction $\theta$. Note that for a \emph{subwavelength} slit array with $d < \lambda$, 
we have $d\sin\theta \leq d <\lambda$, so the \emph{phase difference\/} 
can never reach $2\pi$. This means that the oblique wavelets always tend to 
cancel each other out in the far fields
(destructive interference, similar to the absence of x-ray diffraction at non-Bragg angles). 
Thus, they form evanescent waves near the surface. Mathematically, these
evanescent waves can be expressed as  
${\mathbf E}_m \exp[-(G_m^2-K^2)^{1/2}|z|-iG_m x]$,
where $G_m = 2m\pi /d$ ($m \neq 0$ being integers) and
$K = 2\pi /\lambda$ (note that $|G_m|>K$ for $d<\lambda$) \cite{r9}.

\begin{figure} 
\includegraphics[scale=1.0,angle=0]{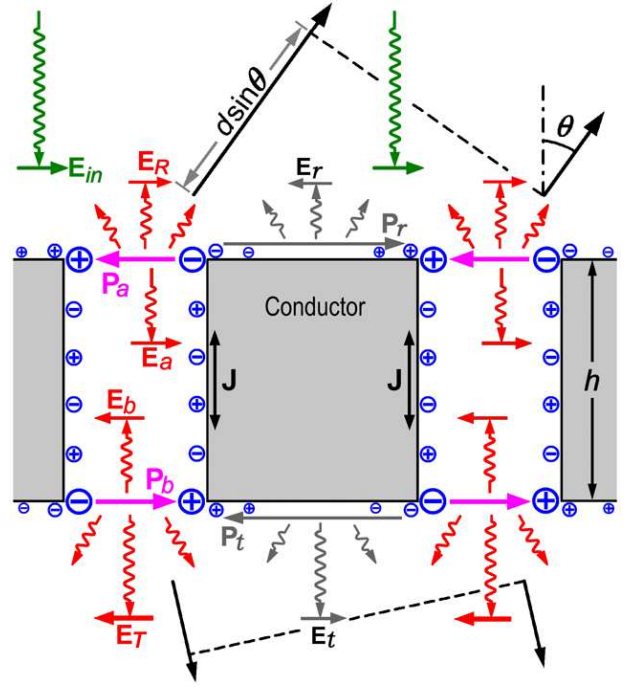}
\caption{Light transmission process 
in the 1D grating. The phases (directions) of the waves and dipoles 
are based on Fig.~2 (and they all have a common oscillating 
factor $e^{i\omega t}$).}
\end{figure}

One may prove that for any wavelength $\lambda>d$, the charge patterns on the upper surface
are \emph{always\/} the same as that in Fig.~2(a) except that the peak heights vary with $\lambda$. 
Therefore, the period of the charge density wave 
always equals the lattice constant $d$ and 
generally does not satisfy the dispersion
relation $k_{sp} = K[\varepsilon_c/(1+\varepsilon_c)]^{1/2}$ of a classical SP (CSP) \cite{r14}
at all
($\varepsilon_c$ the conductor's permittivity and $k_{sp}$ the wavevector of the CSP). 
However, the evanescent waves 
are indeed surface-bound modes with their strengths decaying 
exponentially along $-z$. 
So they are \emph{spoof\/} SPs (SSPs) \cite{r15},
but their formation is the result of \emph{charge oscillation-induced 
light emission and destructive interference\/}. 
The SSP model has been
conceptually proposed by Pendry \emph{et al}.~\cite{r3}
(also see \cite{r16}). Here we provide a concrete 
picture illustrating its origin.

The wavelets propagating along the backward direction ($\theta=0$) 
are always \emph{in phase\/}
(the path difference is zero). Therefore, the back reflected wave is not evanescent but
a propagating
mode. However, since ${\mathbf E}_R$ and ${\mathbf E}_r$ always have opposite directions, 
they tend to offset each other, 
thus reducing the overall back reflection.

In Fig.~3, dipole ${\mathbf P}_a$ also emits a wavelet ${\mathbf E}_a$
inside the slit. 
Similarly, \emph{electrons on the slit walls 
also move in response to the electric field in the slit\/} 
(represented by the current ${\mathbf J}$). Then the charge movement is 
disrupted again at the exit edges, giving rise to another large dipole 
${\mathbf P}_b$. Through this tunneling process, 
the charge patterns  
on the upper surface are duplicated on the lower surface. Oblique wavelets emitted 
from these duplicated light sources 
then form SSPs again below the grating. The
sub-wavelets ${\mathbf E}_T$ and ${\mathbf E}_t$ along the forward
direction form the zero-order transmitted wave. Overall, the grating only emits
two propagating modes, the reflected and transmitted waves, that share the incident energy,
while all the other modes are SSPs.

The oscillating dipole ${\mathbf P}_b$ can give a strong feedback 
to the upper surface by emitting a wavelet ${\mathbf E}_b$ propagating upward. 
If ${\mathbf E}_b$ is \emph{not\/} in phase with 
${\mathbf E}_a$ (and ${\mathbf E}_{in}$) at $z = 0$, it suppresses the strengths of ${\mathbf P}_a$ 
and ${\mathbf E}_R$. 
Then ${\mathbf E}_r$, \emph{which includes specular reflection from the metal surface}, becomes dominant, 
leading to strong backward reflection. The weakened ${\mathbf E}_a$ meanwhile reduces ${\mathbf P}_b$.
However, if ${\mathbf E}_b$
is in phase with ${\mathbf E}_a$ at $z = 0$, it enhances 
${\mathbf P}_a$. 
The enhanced ${\mathbf P}_a$ subsequently strengthens ${\mathbf E}_a$, 
${\mathbf P}_b$, ${\mathbf E}_b$, and so on. 
Then a Fabry-Perot-like resonant state is formed 
with the strengths of all the dipoles and wavelets maximized. Under this condition,
${\mathbf E}_r$ is largely offset by ${\mathbf E}_R$ in the far-field regions,
leading to minimized backward reflection.
Below the grating, wavelet ${\mathbf E}_t$ 
also partly offsets ${\mathbf E}_T$, but $|{\mathbf E}_T|$ 
can be much larger than $|{\mathbf E}_t|$ (unlike ${\mathbf E}_r$, ${\mathbf E}_t$ does not include any specular reflection).
Therefore, when ${\mathbf E}_T$ is  
maximized, it maximizes $T_0$. 
The resonant wavelength is always slightly longer 
than $2h /N$, the ideal Fabry-Perot wavelength,
where $N$ is the resonant number (number of the standing wave nodes). 
Although the Fabry-Perot-like resonance has been recognized
before \cite{r4,r5,r12,r13,r17,r18}, here
we explictily illustrated its origin. Particularly, 
Figure 3 shows that it is dipoles ${\mathbf P}_a$ and ${\mathbf P}_b$ that 
act as the two ``reflecting planes'' required to form a vertical resonator \cite{r13}. 

Note that when $\lambda \leq d$, some of the oblique waves satisfying
$|G_m|\leq K$ become propagating 
diffracted waves (i.e., they are no longer SSP modes).
The sharing of the incident energy by these diffracted waves
significantly reduces $T_0$ (and backward reflection). This explains 
the much lower transmittance in the $\lambda \leq d$ range in Fig.~1. 

Light scattering from a 2D hole array  
has a similar picture. The incident electric field
drives electrons on the upper surface to oscillate. The charge movement 
is blocked at the hole edges, giving rise to oscillating dipoles at the 
entrance openings. 
Thus, the holes act as a 2D array of light sources. For a subwavelength lattice 
with $\max(d, d_2)< \lambda$ (Fig.~4), the phase difference between wavelets emitting from adjacent holes 
again is less than $2\pi$ along any oblique directions. So these wavelets form SSPs above the film except that the
wavelets along $-z$ 
constitute a non-evanescent reflected wave 
(which offsets specular backward reflection). 
Meanwhile, the charge patterns are tunneled onto the lower 
surface due to the charge movement on the hole walls. 
Consequently, a similar set of SSPs and a forward transmitted wave are 
formed below the film. 

However, 2D holes have a different tunneling mechanism.  
Consider the rectangular hole (a unit cell of an array) in Fig.~4. 
Except for the wave distortions near the  
ends,
the electric fields in the hole roughly take the waveguide modes. 
For subwavelength holes, the basic 2D TE$_{1,0}$ modes 
dominate \cite{r3,r11}.
When $\lambda>2L$, 
the modes in the hole are evanescent and approximately take the forms
${\mathbf E}_{a} \propto \sin(\pi y /L) \exp(-\beta z)\hat{\mathbf x}$
and 
${\mathbf E}_{b} \propto \sin(\pi y /L) \exp[-\beta (h-z)]\hat{\mathbf x}$,
where $\beta = \pi(1/L^2-4/\lambda^2)^{1/2}$ \cite{r13}.
${\mathbf E}_a$
drives charges on the walls to move/oscillate, but the charge density decays with decaying
${\mathbf E}_a$ toward $+z$ due to the waveguide restriction. Nevertheless, the sharp discontinuity
of the charge movements at the exit end can still cause significant accumulation of charges there, giving
rise to a relatively large dipole ${\mathbf P}_b$, provided that $h$ is adequately small. 
Since ${\mathbf E}_a$ and ${\mathbf E}_b$ no longer have 
position-dependent phase factors, they are always in phase with each other and 
the incident wave, 
and they always resonate as the feedback ${\mathbf E}_b$ is always constructive.
But the strengths of the dipoles and wavelets decrease with increasing 
$h$ due to the decaying effect.
For fixed $h$ and in the absence of diffraction effects, the transmittance 
would increase monotonically with $\lambda$ decreasing toward $2L$ 
($\beta$ decreasing). However, when $\lambda < \max(d, d_2)$, 
diffracted waves emerge, 
which reduces the zero-order transmittance in the short-wavelength range.
Therefore, strongest transmission must occur for 
$\lambda > \max(d, d_2)$, the non-diffraction range, 
but meanwhile $\lambda$ should still be close to $\max(d, d_2)$ 
so that the damping coefficient $\beta$ is sufficiently small.

\begin{figure} 
\includegraphics[scale=1.0,angle=0]{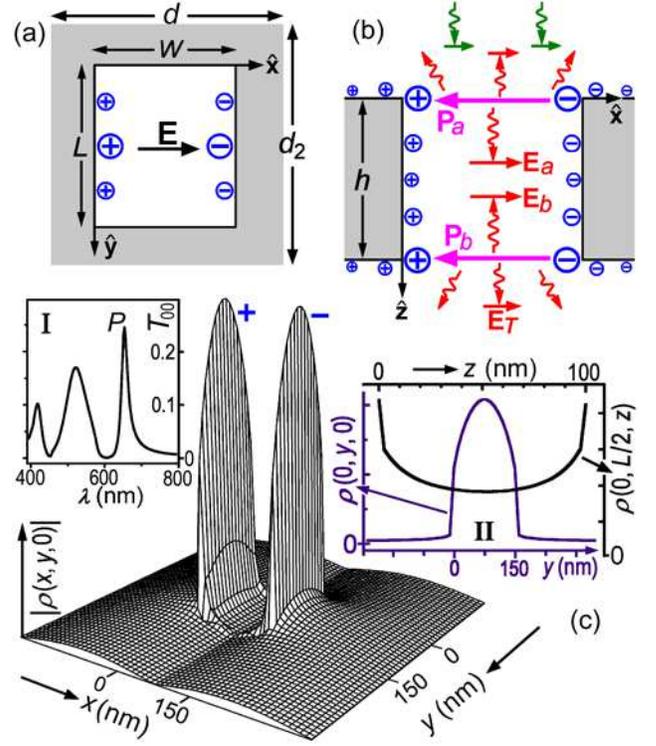}
\caption{Light transmission through 2D 
hole arrays. (a) The unit cell of 2D lattice viewed along $z$. 
(b) Side view. Here the dipoles similar to those denoted by ${\mathbf P}_r$ in Fig.~3 are omitted. 
(c) Charge densities $|\rho(x,y,0)|$ of a silver 
hole array: $h = 100$, $W = L = 150$, $d = d_2 = 600$ nm. 
$\lambda = 673$ nm, corresponding to peak $P$ in 
the zero-order transmittance ($T_{00}$) spectrum in Inset I. 
Inset II: Profiles $\rho(0,y,0)$ and 
$\rho(0,L/2,z)$ with constant phases at $\lambda = 673$ nm.}
\end{figure}

This picture agrees very well with the measured transmission spectra 
in the literature, 
and our finite-difference time-domain (FDTD) computations 
have unambiguously proved it. Figure 4(c) shows the 
calculated charge distribution $|\rho(x,y,0)|$
on a free-standing silver film with a hole array,
where the charges again exist mainly on the two walls $x=x_w$ (perpendicular 
to the charge movement direction).
The residual charges in other regions are surface charges
and they disappear for $0<z<h$. (Here note that FDTD
gives the overall charge densities.)
The shape of the 
$\rho(0, y, z)$ [$\equiv -\rho(W,y,z) \sim \sin(\pi y /L)$] profile 
is independent of $z$, but the maximum density
$\rho(0, L/2, z)$ changes with $z$ in
inset II. Here our calculations indeed reveal the two large dipoles at 
the two openings of the hole (also see \cite{r19}
for the single-hole case). 

The above illustrations may indicate a general Thomson scattering
mechanism of metallic nanostructures 
similar to x-ray scattering by oscillating electrons. 
That is, free electrons on a non-planar conducting surface 
cannot move homogeneously in response to the incident wave, 
thus forming inhomogeneous charge distributions.  
The oscillating charges then emit wavelets, and the interference
between the wavelets may give rise to
anomalous light scattering. The 
basic requirement here is free electrons, so this mechanism
can explain scattering
from various conducting nanostructures, including
perfect conductors and conductors with Re$(\varepsilon_c)>0$
(see the TM$'$ curve in Fig.~1 and the 
experiments and simulations of tungsten hole arrays with positive
Re$(\varepsilon_c)$ in \cite{r8,r20}). It is also applicable to non-periodic
structures. For example, Figure 3 indicates that an \emph{isolated\/} slit (or hole)
can still be a single light source emitting \emph{divergent\/} light. If the slit is surrounded by grooves
(in which the cavity resonance may be different though),
the grooves provide additional light sources that suppress the oblique wavelets (forming SSPs).  
Meanwhile, wavelets emitted from the grooves along the backward direction offset specular reflection. 
Thus, transmission
is enhanced. If grooves also exist on the exit surface, they again suppress oblique wavelets so that a  
collimated zero-order transmitted beam can be achieved \cite{r2}. 
As another example, it is known that
when a nanowire is illuminated by a TM wave at one end, light can be ``transferred'' 
to the other end \cite{r21,r22}. 
The common explanation is that light is transferred by CSPs on the wire, but the dispersion trend 
in \cite{r21} is quite different from that of CSPs.
Based on our mechanism, the incident wave causes inhomogeneous charges 
at the illuminated end that tend to propagate away due to the charge-netural tendency. 
This is similar to
normal electricitiy transmission over metallic wires. The charge movement is then discontinued at the other end, leading to charge accumulation and oscillation 
that emit new light. The electrons can be bounced back, resulting in a Fabry-Perot-like charge pattern
(that obviously do not need to obey the CSP principle).
Here high conductivity can enhance the transfer efficiency, which is opposite to the CSP prediction
that (nearly) perfect conductors do not support CSPs.
Overall, our mechanism reveals the fundamental of metallic nanostructure optics. 
 
X.R.H is grateful to A. T. Macrander and S. K. Gray 
for invaluable suggestions and discussions.
This work was supported by the U.S. Department of Energy, 
Office of Science, Office of Basic Energy Sciences, under
Contract No.~DE-AC02-06CH11357, 
and by the National Nature Science Foundation of China 
(10625417, 50672035, 10021001) and the Ministry of Science and 
Technology of China (the State Key Program for Basic Research, 
2004CB619005, 2006CB921804).


\begin{thebibliography}{}
\bibitem{r15} The period of a CSP also satisfies 
$2\pi/\mbox{Re}(k_{sp}) < \lambda$ [for finite $\varepsilon_c$ 
with $\mbox{Re}(\varepsilon_c)<0$]. A sinusoidal CSP on a planar 
surface is a true surface mode since along the normal direction,
the two sub-wavelets emitted from a period are symmetric so that they 
are always cancelled out. CSP excitation
depends on the intrinsic dielectric properties of the metal while SSPs 
corresponds to a geometrical effect.
\end{thebibliography}
\end{document}